\documentclass[12pt]{iopart}

\usepackage{amsfonts}
\usepackage{amsthm}
\usepackage{amssymb}
\usepackage{bbm}
\usepackage{graphicx,hyperref}
\usepackage{epsfig} 
\usepackage{verbatim}
\newcommand{\defeq}{\mathrel{\mathop:}=}
\begin{document}
\title{Kinetic theory of quantum and classical Toda lattices}
\author{Vir B. Bulchandani$^{1}$, Xiangyu Cao$^{1}$, Joel E. Moore$^{1,2}$ }
\address{$^1$ Department of Physics, University of California, Berkeley, Berkeley CA 94720, USA}
\address{$^2$ Materials Science Division, Lawrence Berkeley National Laboratory, Berkeley CA 94720, USA}
\begin{abstract}
 We consider the kinetic theory of the quantum and classical Toda lattice models. A kinetic equation of Bethe-Boltzmann type is derived for the distribution function of conserved quasiparticles. Near the classical limit, we show that the kinetic theory depends smoothly on Planck's constant, and explicitly characterise the leading quantum corrections to classical behaviour. The classical kinetic theory is compared with direct numerical simulations and shows excellent agreement. Finally, we connect the Bethe-Boltzmann approach with the classical inverse scattering method, by identifying conserved quasiparticles with the spectrum of the Lax matrix.
\end{abstract}

\section{Introduction}  

The derivation of hydrodynamics from microscopic many-particle dynamics is an enduringly difficult mathematical problem. Despite the technical hurdles involved, rigorous progress has been made for some simple classical many-body systems~\cite{SpohnBook}.  However, beyond the relatively tractable case of weakly interacting Bose gases~\cite{GPE}, the hydrodynamics of quantum many-body systems remains poorly understood.

Recently, a promising new avenue of inquiry has opened up with the development of ``generalized hydrodynamics'' of quantum integrable systems in one dimension~\cite{Castro-Alvaredo2016,Bertini2016}, which has been accompanied by rigorous results on the local equilibrium dynamics of one-dimensional conformal field theories~\cite{Langmann2017,Gawedzki2018}. The Bethe-Boltzmann kinetic equation for quasiparticle densities, which captures some essential features of generalized hydrodynamics with an infinite set of conserved quantities and Euler equations, has yet to be derived rigorously, but there is a widespread belief in the field, supported by a wealth of numerical and analytical tests~\cite{Castro-Alvaredo2016,Bertini2016,Ilievski2017,BVKM1,BVKM2,Doyon20172,Ilievski20172,Piroli2017,Fagotti2017,Bastianello2018} (and a recent experiment~\cite{Schemmer2018}), that this equation describes the ballistic part of local equilibrium dynamics in such models exactly.

The kinetic equation can be understood as a generalization of earlier kinetic theories of classical soliton gases~\cite{Percus69,Zakharov71,Boldrighini83,El2003,El2005,Doyon2017,Doyon2018,Bulchandani2017}, with classical phase-shifts replaced by the quantum mechanical phase-shifts between Bethe quasiparticles.  The concept that the dynamics of Bethe quasiparticles could capture far-from-equilibrium transport in quantum integrable spin chains was discussed early on~\cite{zotos}.  Given that the kinetic approach amounts to describing quantum dynamics by an essentially classical equation, a natural question that arises is the following: how does a kinetic theory of quantum solitons differ from that of classical solitons? The integrable Toda lattice, whose thermodynamic Bethe Ansatz (TBA) is understood both classically and quantum mechanically~\cite{Hader1986,GrunerBauer1988,Siddharthan1996}, provides an ideal setting in which to analyse this important question.

In the present work we derive Bethe-Boltzmann kinetic equations for the quantum and classical Toda lattices (Section \ref{sec:BB}). Hydrodynamic predictions obtained for the classical Toda lattice are compared against microscopic numerical simulations of ``two-reservoir'' initial conditions  and show excellent quantitative agreement (Section \ref{sec:num_studies}). Our kinetic equation for the classical Toda lattice can be understood as the semiclassical limit of the kinetic theory of the quantum system, allowing for a direct, quantitative analysis of the effect of introducing a small and non-zero $\hbar$ in the Bethe-Boltzmann equation.  Hence the Toda problem illustrates, in a single model with variable quantum parameter $\hbar$, how the same type of kinetic equation describes the hydrodynamic limit of quantum and classical integrable systems.

Finally, we make a connection between equilibrium quasiparticle distribution functions of the classical Toda lattice and classical inverse scattering theory, interpreting the local distribution functions of quasiparticle density and current directly in terms of the spectrum of the Lax matrix. This ``Bethe-Lax correspondence'' turns predictions of the kinetic theory to mathematical statements about tridiagonal random matrix ensembles. They are convincingly supported by our numerical results, which indicate that the conjectured equations of state for the quasiparticle density and current distribution functions in Section \ref{sec:BB} hold exactly for equilibrium states. The Bethe-Lax correspondence offers a new qualitative insight into the origin of this powerful hydrodynamic formalism. In the classical Toda case, an intriguing outcome would be a rigorous derivation of our equation of state for the quasiparticle \emph{currents} from Dumitriu-Edelman $\beta$-random matrix theory~\cite{rmt}, as was very recently achieved for the quasiparticle density of states~\cite{spohn}.
We note that the hydrodynamics of classical Toda lattice is treated in an extensive recent work~\cite{doyon} using a different approach.

\section{Kinetic equations: quantum and classical}
\label{sec:BB}  

\subsection{Quantum TBA and kinetic theory}
The quantum Toda Hamiltonian can be defined as
\begin{equation}
H = \sum_n -\frac{\hbar^2}{2m}\partial_n^2 + V(x_{n+1}-x_n) \,,\, 
V(r) = \frac{m\omega^2}{\gamma^2} e^{-\gamma r} \, . \label{eq:def-toda}
\end{equation}
Here $\gamma$ quantifies the anharmonicity of the potential and $\omega$ is the harmonic frequency as $\gamma \to 0$. Using Sutherland's technique of ``asymptotic Bethe ansatz'', one can deduce the equations of thermodynamic Bethe ansatz~\cite{Hader1986} for this model. At temperature $T$ and chemical potential $\mu$, the Yang-Yang equation for excitation energies is given by
\begin{equation}
\epsilon_k = \frac{\hbar^2k^2}{2m} - \mu + \frac{1}{\beta}\int_{-\infty}^{\infty} dk' \, K_{k,k'}\ln{(1+e^{-\beta\epsilon_{k'}})}\, , \label{eq:YY1}
\end{equation}
while the total density of states $\rho^t_k$ satisfies
\begin{equation}
\rho_k^t = \frac{1}{2\pi} -\int_{-\infty}^\infty dk' \, K_{k,k'}\theta_{k'}\rho^t_{k'} \, ,
\label{eq:bethe-quantum}
\end{equation}
where $\theta_k = (1+e^{\beta \epsilon_k})^{-1}$ denotes the  Fermi factor. The differential phase-shift is given explicitly by
\begin{equation}
K_{k,k'} = \frac{1}{2\pi \gamma}\left[2 \ln{C} - \psi(1-i(k-k')/\gamma) - \psi(1+i(k-k')/\gamma)\right]\, . \label{eq:K-qu}
\end{equation}
Here, the dimensionless parameter $C=\frac{m\omega}{\hbar \gamma^2}$ and $\psi(x)$ denotes the digamma function. Abstracting to general local equilibrium states, with arbitrary Fermi factors $\theta_k : \mathbb{R} \to [0,1]$, we find that the derivatives of quasiparticle energies and momenta satisfy
\begin{eqnarray}
\label{eq:q_dr}
\nonumber \epsilon'_k +\int_{-\infty}^\infty dk' \, K_{k,k'}\theta_{k'}\epsilon'_{k'} &= \hbar^2k/m \\
{p^{dr}}'_k +\int_{-\infty}^\infty dk' \, K_{k,k'}\theta_{k'}{p^{dr}}'_{k'} &= \hbar \, .
\end{eqnarray}

From these formulae, one can write down a ``Bethe-Boltzmann'' equation, modelling the ballistic part of locally equilibrated time-evolution in the quantum Toda lattice. The resulting formalism is identical to that for the Lieb-Liniger gas~\cite{Castro-Alvaredo2016,Bertini2016}, and the initial value problem for smoothly varying temperature and chemical potential profiles $\beta(x)$ and $\mu(x)$ can be conveniently summarized in terms of local Fermi factors $\theta_k(x,t)$, which evolve under the system of equations
\begin{eqnarray}
\label{eq:q_adv}
\partial_t \theta_k + v_k[\theta]\partial_x \theta_k = 0, \, \, v_k[\theta] = \frac{(\hat{1} + \hat{K}\hat{\theta})^{-1}[\hbar k'/m]_k}{(\hat{1} + \hat{K}\hat{\theta})^{-1}[1]_k}\, ,
\end{eqnarray}
with initial data determined locally in terms of $\{\beta(x),\mu(x)\}$ via thermodynamic Bethe Ansatz. Here, the notation $\hat{O}$ indicates the integral operator corresponding to a given integral kernel $O_{k,k'}$. In particular, $(\hat{1}+\hat{K}\hat{\theta})^{-1}$ stands for the inverse of the integral operator with kernel  $\delta(k-k')+K_{k,k'}\theta_{k'}$, that appears in Eq. ~\ref{eq:q_dr}. Thus $(\hat{1} + \hat{K}\hat{\theta})^{-1}[f]_k \defeq \int dk' \, U_{k,k'}f_{k'}$, with $U_{k,k'}$ determined by the integral equation
\begin{equation}
U_{k,k'} + \int dk'' K_{k,k''}\theta_{k''}U_{k'',k'} = \delta(k-k').
\end{equation}

\subsection{The semiclassical limit of TBA}

We now turn to the semiclassical limit $\hbar \to 0$ of the quantum TBA and hydrodynamics outlined above, following closely the analysis of Ref.~\cite{GrunerBauer1988}. We proceed in two stages: we first express the quantum TBA equations in terms of classical momentum $p = \hbar k$, before taking the semiclassical limit as $\hbar \to 0$. In order to guarantee a finite free energy in the semiclassical limit~\cite{GrunerBauer1988}, we define classical quasiparticle energies and classical chemical potentials, as
\begin{eqnarray}
\label{eq:chempot}
\epsilon^{cl}_p = \epsilon_p + \frac{1}{\beta} \ln{\beta\hbar\omega} \,,\,
\mu^{cl} = \mu- \frac{1}{\beta} \ln{\beta\hbar\omega}\, .
\end{eqnarray}
The Yang-Yang equation for $\epsilon^{cl}_p$ reads
\begin{equation}
\epsilon^{cl}_p = \frac{p^2}{2m} - \mu^{cl} + \frac{1}{\beta \hbar}\int_{-\infty}^{\infty} dp' \, K_{p/\hbar,p'/\hbar}\ln{(1+\beta \hbar \omega e^{-\beta\epsilon^{cl}_{p'}})} \, ,
\label{eq:qc_yy}
\end{equation}
the Fermi factor is given by
\begin{equation}
\label{eq:qc_theta}
\theta_p = \frac{\beta \hbar \omega e^{-\beta^{cl}_p}}{1+\beta \hbar\omega e^{-\beta \epsilon^{cl}_p}} \, ,
\end{equation}
and the dressing equations read
\begin{eqnarray}
\label{eq:qc_dress1}
{\epsilon^{cl}}'_p +\int_{-\infty}^\infty dp' \, K_{p/\hbar,p'/\hbar}(\theta_{p'}/\hbar){\epsilon^{cl}}'_{p'} &= p/m \, , \\
\label{eq:qc_dress2}
{p^{dr}}'_p +\int_{-\infty}^\infty dp' K_{p/\hbar,p'/\hbar}(\theta_{p'}/\hbar){p^{dr}}'_{p'} &= 1 \, .
\end{eqnarray}
The rescaled density of states, defined by $\widetilde{\rho}_p^t dp = \rho^t_k dk$, satisfies $2\pi \hbar \widetilde{\rho}^t_p = {p^{dr}}'_p$. Note that equations Eq. \ref{eq:chempot} through \ref{eq:qc_dress2} are merely a rewriting of quantum TBA, and are exact for $\hbar>0$. Later, they will allow us to quantify the quantum corrections to classical hydrodynamics. Let us first consider passing to the semiclassical limit $\hbar \to 0$ of quantum TBA. 

In this limit, the phase-shift is replaced by its leading asymptotic,
\begin{eqnarray}
\label{eq:K-cl}
K^{cl}_{p,p'} &= \lim_{\hbar \to 0} K_{p/\hbar,p'/\hbar} = -\frac{1}{\pi\gamma}\left(\ln\left|\frac{\gamma}{m\omega}\right| + \ln{|p-p'|}\right) \,,
\end{eqnarray}
the Yang-Yang equation becomes
\begin{eqnarray}
\label{eq:class_yy}
\epsilon^{cl}_p &= \frac{p^2}{2m} - \mu + \int_{-\infty}^{\infty} dp' \, K^{cl}_{p,p'} \omega e^{-\beta\epsilon^{cl}_{p'}}\, ,
\end{eqnarray}
and the effective Fermi factor appearing in the dressing equations is given by
\begin{eqnarray}
\label{eq:class_ff}
\widetilde{\theta}_p = \lim_{\hbar \to 0} \theta_p/\hbar = \beta \omega e^{-\beta \epsilon^{cl}_p}\, .
\end{eqnarray}
In particular, the density of occupied states reads
\begin{eqnarray}
\label{eq:class_dos}
\widetilde{\rho}_p = \lim_{\hbar \to 0} \widetilde{\rho}^t_p \theta_p = \widetilde{\theta}_p(1+\hat{K}_{cl}\hat{\widetilde{\theta}})^{-1}(1/2\pi)_p \, ,
\end{eqnarray}
and from this expression, the thermal expectation values of conserved charges of the classical Toda lattice, $\langle Q_n \rangle_{\rho} = \int dp \, \widetilde{\rho}_p p^n$, may be evaluated. The relation Eq.~\ref{eq:class_dos} is verified numerically in Section \ref{sec:num_studies}, by direct comparison with spectrum of the Lax matrix.

\subsection{Classical kinetic theory}

For any non-zero $\hbar$, the quasiparticle distribution function of the full quantum TBA includes both ``phonon'' (particle-like) and ``soliton'' (hole-like) degrees of freedom. However, in the semiclassical limit at non-zero temperature, the soliton degrees of freedom become infinitely heavy and do not contribute to the thermodynamics of the model~\cite{GrunerBauer1988}. Thus for the purposes of considering local equilibrium dynamics in the Toda lattice, it suffices to restrict attention to phononic degrees of freedom; this will be justified more carefully below. The phonon dispersion relation is given by~\cite{GrunerBauer1988}
\begin{eqnarray}
\Delta E^{ph}_p &= \hbar \omega e^{-\beta \epsilon^{cl}_p}, \, \, \Delta P^{ph}_p = 2\pi \hbar \int_{p}^\infty dp' \, \widetilde{\rho}_{p'},
\end{eqnarray}
and implies an interaction-dressed phonon group velocity
\begin{equation}
v_p = \frac{\Delta {E^{ph}}'_p}{\Delta {P^{ph}}'_p} = \frac{{\epsilon^{cl}_p}'}{2\pi \hbar \widetilde\rho_p^t},
\end{equation}
for a state with occupied density of states $\widetilde{\rho}_p$. From this expression, we can write down a Bethe-Boltzmann equation for the local phonon density of states, given by
\begin{equation}
\label{eq:class_BB}
\partial_t \widetilde{\rho}_p(x,t) + \partial_x (\widetilde{\rho}_p(x,t) v_p[\widetilde{\rho}](x,t)) = 0.
\end{equation}
As usual, this may be expressed in advection form as the system
\begin{eqnarray}
\label{eq:class_adv}
\partial_t \widetilde{\theta}_p + v_p[\widetilde{\theta}]\partial_x \widetilde{\theta}_p &= 0, \, v_p[\widetilde{\theta}] &= \frac{(\hat{1}+\hat{K}^{cl} \hat{\widetilde{\theta}})^{-1}[p'/m]_p}{(\hat{1}+\hat{K}^{cl} \hat{\widetilde{\theta}})^{-1}[1]_p},
\end{eqnarray}
with initial data determined in terms of initial temperature and (classical) chemical potential via the classical TBA equations Eqs. \ref{eq:class_yy} and \ref{eq:class_ff}. Solutions to this system can be compared directly with classical numerical simulations of Toda lattice dynamics. Such a comparison is provided in Section \ref{sec:num_studies}.

While the above discussion was physically motivated, notice that the classical hydrodynamic equation Eq. \ref{eq:class_adv} could have been obtained directly from the quantum hydrodynamic equations Eq. \ref{eq:q_adv} by taking the semiclassical limit described above (the possibility of obtaining classical hydrodynamics as a semiclassical limit was noted for the sinh-Gordon model in Ref.~\cite{Bastianello2018}). Since the quantum distribution function $\rho_k$ does not distinguish between phonon and soliton degrees of freedom, this indicates that solitons can indeed be neglected at the level of classical hydrodynamics. This conclusion is supported by the numerical results of Section \ref{sec:num_studies}.

\subsection{Quantum corrections to classical kinetic theory}

\begin{figure}[h]
    \centering
    \includegraphics[width=0.65\linewidth]{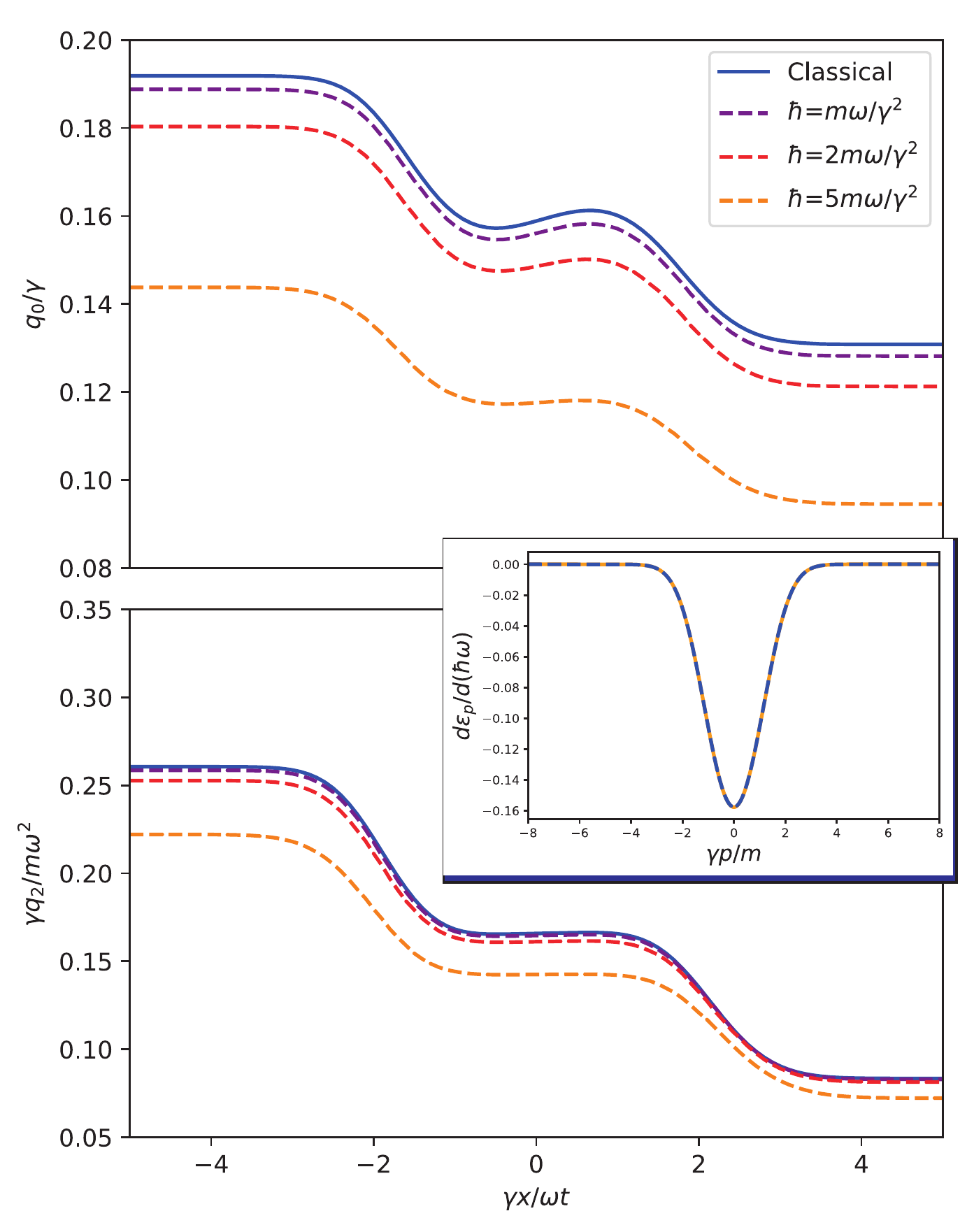}
    \caption{Main plots: Tuning $\hbar$ in the two-reservoir steady state of the Toda lattice predicted by hydrodynamics, with initial conditions $\beta_L = 0.5, \, \mu_L = -1$ and $\beta_R = 1, \,\mu_R = -1$. Observables plotted are particle density, $q_0(x,t)$ (\textit{top}) and energy density $q_2(x,t)$ (\textit{bottom}), in units $\{m,\gamma^{-1},\omega^{-1}\}$ set by the microscopic model. Inset: Analytical prediction (Eq.~\ref{eq:e_correction}) for leading quantum correction of the dressed energy $ \epsilon^{cl} $ (dashed), versus result obtained from directly solving Eq. \ref{eq:qc_yy} by numerical iteration for $0< \hbar \ll 1$.}
    \label{fig:q_corr}
\end{figure}

We now consider the quantum corrections to the kinetic theory of the classical Toda lattice obtained above. The quantum TBA equations written in terms of classical momenta, Eqs. \ref{eq:qc_yy} - \ref{eq:qc_dress2}, allow for a direct comparison between quantum and classical results, yielding predictions at a given temperature and (classical) chemical potential that tend smoothly to the semiclassical results as $\hbar \to 0$. This is shown in Fig. \ref{fig:q_corr} for the two-reservoir non-equilibrium steady state studied in Section \ref{sec:num_studies}. As the dimensionless scale $m \omega / \hbar \gamma^2 $ falls below $1$, deviations from the classical limit become apparent. The main effect of increasing the relative magnitude of $\hbar$ appears to be a reduction in particle density in each reservoir, with a concomitant decrease in energy density.

Let us now quantify these corrections analytically. It turns out that the main subtlety arises from the quantum correction to the phase shift, which we focus on at first. Then, as an application, we calculate explicitly the first-order correction to the dressed energy. Our results can be straightforwardly extended to higher-order and other quantities.

The quantum correction to the phase shift is naturally defined as the difference between quantum and classical phase shifts. By Eqs.~\ref{eq:K-qu} and \ref{eq:K-cl}, we have:
\begin{eqnarray}
    \delta K_{p,p'} = K_{p/\hbar,p'/\hbar} - K_{p,p'}^{cl} \nonumber\\
    = \frac{1}{\pi\gamma}\left(\ln {|p-p'|/\gamma \hbar} - \mathrm{Re}\{\psi(1+i  |p-p'|/\gamma\hbar)\}\right)\,. \label{eq:dK}
\end{eqnarray}   
We wish to analyze the semiclassical expansion of this quantity about $\hbar = 0$. The final result will be given in Eq.~\ref{eq:K_expansion}. However, we find it instructive to first consider a \textit{wrong} approach. Indeed, one might be tempted to perform an asymptotic expansion in the limit $\Delta k := \Delta p / \hbar \to \infty$, which corresponds to the limit $\hbar \to 0$ with fixed $\Delta p \neq 0$. Using the asymptotic expansion~\cite{ASbook}
\begin{eqnarray}
\mathrm{Re}\{\psi(1+ik)\} \sim \ln k + \sum_{n=1}^\infty \frac{(-1)^{n-1}}{2n k^{2n}}B_{2n}, \, k \to \infty, \label{eq:bad-expansion}
\end{eqnarray}
of the digamma function ($B_{2n}$ denote the Bernoulli numbers), one would conclude that leading quantum correction is of order $\hbar^2$:
\begin{eqnarray}
\delta K_{p,p'} = -\hbar^2 \frac{\gamma}{12\pi}\frac{1}{|p-p'|^2} \,,\, \mathrm{for \; fixed \;} 
|p-p'|>0 \,.
\end{eqnarray}
However, the correction kernel on the right hand side is pathological because of the divergence at $p-p' \to 0$. In fact, the expansion \ref{eq:bad-expansion} gives diverging kernels at \emph{all} orders $\hbar^{2n}$, and must be unphysical. This indicates that the quantum correction comes essentially from static (``soft'') scattering $p-p' \to 0$ (more precisely, $|p-p'|\sim \hbar \gamma$).   
 
This is indeed the case. In fact, we claim that $\delta K_{p,p'}$ is concentrated at $p-p'=0$ at any order of $\hbar$. More precisely, 
\begin{eqnarray}
    \delta K_{p,p'} = \sum_{n=0}^{\infty}  \frac{\kappa_{2n}}{(2n)!} \delta^{(2n)}(p-p') \gamma^{2n} \hbar^{2n+1} \,,\,  \label{eq:K_expansion} \\ 
   \mathrm{where}\, \kappa_{2n} =  \frac{1}{\pi } \int_{-\infty}^{\infty} d u \, u^{2n}\left[ \ln |u|  - \mathrm{Re}\{\psi(1+i u)\} \right] \,,  \nonumber
\end{eqnarray}
and $\delta^{(s)}$ is the $s$-th derivative of $\delta$. Curiously, the semiclassical expansion of the phase shift has only odd positive powers of $\hbar$, although the phase shift is an even function of $(p-p')/\hbar$. In particular, the leading order is of order $\hbar$ (its coefficient can be explicitly evaluated):
\begin{equation}
    \delta K_{p,p'} = \hbar \kappa_0 \delta(p-p') + \mathcal{O}(\hbar^2)   \,,\, \kappa_0 = -\frac12 \,.\label{eq:delta}
\end{equation}

To show Eq.~\ref{eq:K_expansion}, we apply the kernel $ \delta K_{p,p'}$ to an arbitrary smooth function $f_p$ of $p$, write the integral in variables $k = p/\hbar , k' = p' / \hbar$ at the quantum scale, and then Taylor expand $f_{p'}$ around $p$:
\begin{eqnarray}
    \int_{-\infty}^{\infty}  dp' \delta K_{p,p'} f_{p'} & = 
 \hbar    \int_{-\infty}^{\infty}  dk' \delta K_{k,k'} f_{k'\hbar} \nonumber \\
 & = \hbar   \int_{-\infty}^{\infty}  dk' \delta K_{k,k'} \sum_{n=0}^{\infty} \frac{1}{n!} f^{(n)}_{k\hbar} \hbar^n (k'-k)^n   \label{eq:taylorexpand}
\end{eqnarray}
where $f^{(n)} = d^n f_p/ dp^n$. Such an expansion is physically justified because $ f_{k\hbar} $ is a very slowly varying function of $k$. At each order of $\hbar$, the integral over $k'$ is decoupled from $f_p$ by a change of variable ($u = (k'-k)\gamma$)
\begin{equation}
    \int_{-\infty}^{\infty}  dk' (k'-k)^{n} \delta K_{k,k'} 
    = \int_{-\infty}^{\infty}\frac{ d u}{\pi} \left[ \ln |u|  - \mathrm{Re} \psi(1+i u) \right] u^{n} \gamma^{n}  = \kappa_{n} \gamma^{n} \,,  \label{eq:const_integral}
\end{equation}
where we used Eq.~\ref{eq:dK}. Combining Eqs.~\ref{eq:taylorexpand} and \ref{eq:const_integral}, we obtain
\begin{equation}
    \int_{-\infty}^{\infty}  dp' \delta K_{p,p'} f_{p'} =
   \hbar  \sum_{n=0}^{\infty} \frac{1}{n!} f^{(n)}_{p} (\hbar \gamma)^n  \kappa_{n} \,,
\end{equation}
which is equivalent to Eq.~\ref{eq:delta}, since Eq.~\ref{eq:const_integral} vanishes for odd $n$ by parity.


Eq.~\ref{eq:K_expansion} allows the semiclassical expansion of the kinetic theory to any order. To illustrate this, we compute the leading quantum correction to the dressed energy. To this end, we expand it as follows:
\begin{eqnarray}
\epsilon^{cl} = {\epsilon^{cl}}^{(0)} + \delta \epsilon^{cl} + \mathcal{O}(\hbar^2) \,.
\end{eqnarray}
It follows from the Yang-Yang equation in the form Eq. \ref{eq:qc_yy} that the perturbation $\delta \epsilon^{cl}_p$ satisfies the integral equation
\begin{eqnarray}
 & \delta \epsilon^{cl}_p + \int_{-\infty}^{\infty} dp' K^{cl}_{p,p'} \widetilde{\theta}^{(0)}_p\delta \epsilon^{cl}_{p'} \nonumber \\
= & -\frac{1}{2}(\beta \hbar\omega)\int_{-\infty}^{\infty} dp' K^{cl}_{p,p'} \omega e^{-2\beta {\epsilon^{cl}}^{(0)}_{p'}} + \int_{-\infty}^{\infty}  dp' \delta K_{p,p'} \omega 
e^{- \beta {\epsilon^{cl}}^{(0)}_{p'}} \label{eq:YY_correction}
\end{eqnarray}
up to $O(\hbar)$ corrections. Applying Eq.~\ref{eq:delta}, we obtain the integral equation that determines the leading quantum correction to $\epsilon^{cl}_p$:
\begin{eqnarray}
    \delta \epsilon^{cl}_p + \int_{-\infty}^{\infty} dp' K^{cl}_{p,p'} \delta \epsilon^{cl}_{p'}  \nonumber \\
    =    -\frac{\hbar \beta \omega}{2} \int_{-\infty}^{\infty} dp' K^{cl}_{p,p'} \omega e^{-2\beta {\epsilon^{cl}}^{(0)}_{p'}} - \frac12 \hbar \omega e^{-\beta {\epsilon^{cl}}^{(0)}_{p}}  + O(\hbar^2) \,. \label{eq:e_correction}
\end{eqnarray}
This expression is exact and can be verified numerically, see Fig.~\ref{fig:q_corr} (inset) for an example. Observe that in Eq.~\ref{eq:e_correction}, the only kernel that relates unequal momenta $p - p' > 0$ is the classical one $K_{p,p'}^{cl}$. This is a direct consequence of Eq.~\ref{eq:K_expansion} and holds at any order and for all other quantities (e.g., the velocity functional). We also note that the first term of the semiclassical expansion Eq. \ref{eq:e_correction} is controlled by the dimensionless parameter $\lambda = \beta \hbar \omega$, so is valid at fixed nonzero temperature. 

\section{Numerical studies}\label{sec:num_studies}
In this section, we numerically test the classical kinetic theory in two ways.  In Section~\ref{sec:twores}, we test the kinetic theory directly in a non-equilibrium two-reservoir setting. In Section~\ref{sec:lax}, we connect the quasiparticles in kinetic theory to the spectrum of the Lax matrix. For convenience, we set $\gamma = m = \omega = 1$ throughout this section.

\subsection{Out-of-equilibrium dynamics}\label{sec:twores}
We now compare predictions from the classical kinetic theory with microscopic numerical simulations of the classical Toda lattice. While performing direct microscopic simulations on the quantum Toda lattice is not feasible, previous comparison of results from integrability to microscopic DMRG simulations on the XXZ chain~\cite{BVKM2} suggest that the quantum kinetic equation is comparably accurate in the hydrodynamic limit.

For the sake of simplicity, we focus on time-evolution from ``two-reservoir'' initial conditions, for which the solution to the hydrodynamic equations \ref{eq:class_adv} is well-known by now~\cite{Castro-Alvaredo2016,Bertini2016}. To obtain hydrodynamic predictions, the reservoir occupation numbers $\widetilde{\theta}^{L/R}_p$ are computed from the classical TBA equations Eq. \ref{eq:class_yy}, from which the occupation numbers $\widetilde{\theta}_p(x,t)$ for the hydrodynamic steady-state are obtained by solving the system
\begin{eqnarray}
\label{eq:NESS}
\widetilde{\theta}_p (x,t) = \widetilde{\theta}^L_p + (\widetilde{\theta}^R_p - \widetilde{\theta}^L_p)\Theta(x-v_p[\widetilde{\theta}(x,t)]t)
\end{eqnarray}
by numerical iteration, as usual.

\begin{figure}[h]
\centering\includegraphics[width=0.7\linewidth]{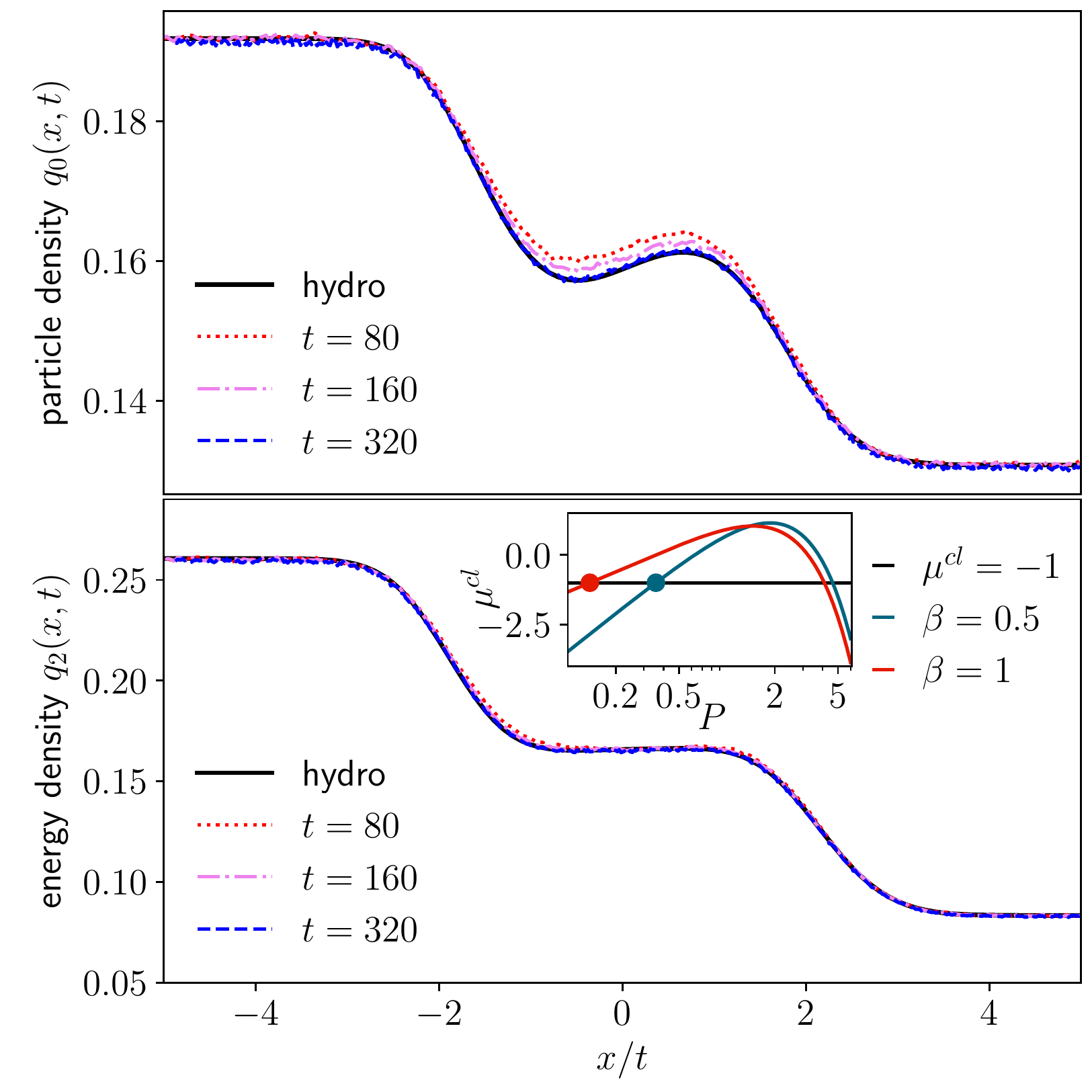}
\caption{Time-evolution of particle density $q_0(x,t)$ and energy density $q_2(x,t)$ from two-reservoir initial conditions with $\beta_L = 0.5, \, \mu_L = -1$ (pressure $P=0.357\dots$) and $\beta_R = 1, \,\mu_R = -1$ ($P = 0.137\dots$). The dashed lines are obtained from microscopic numerical simulations as described in the main text, and the solid line denotes the hydrodynamic prediction. The numerical data is obtained from averaging $ 10^8$ independent samples on a Toda lattice with $2N=1200$ sites and open boundary conditions. See text around Eq.~\ref{eq:Z} for numerical methods. \textit{Inset}: Chemical potential as a function of pressure with temperatures $\beta_L=0.5$, $\beta_R = 1$. The dots correspond to the selected pressures.}
\label{Fig1}
\end{figure}
We then obtain, via Eq. \ref{eq:class_dos}, profiles of particle density and energy density
\begin{equation}
q_0(x,t) = \int dp \, \widetilde{\rho}_p(x,t) \,,\, q_2(x,t) =  \frac{1}{2}\int dp \, p^2 \widetilde{\rho}_p(x,t) \,.
\end{equation}
These predictions can be compared to direct numerical simulations. The equations of motion are integrated using the Flaschka variables~\cite{Flaschka}, see Eq.~\ref{eq:anbn} below. The thermal equilibrium of each reservoir can be sampled directly, since the momenta and positions have the following factorized distribution:
\begin{equation}
    P\left(\{r_n, p_n \}\right) = \frac1Z \prod_n e^{-\beta \frac{p_n^2}2} \exp\left( -\beta e^{-r_n} - \beta P  r_n \right) \,,\, r_{n} = x_{n+1} - x_n \,, \label{eq:Z}
\end{equation}
where $Z$ is the partition function and $P>0$ is the pressure. $\beta$ and $P$ determine the classical thermal state, of which the chemical potential and average stretch are
    \begin{equation}
       \mu^{cl} =  -\frac{1}\beta\left( \ln \Gamma(\beta P) - \ln \sqrt{2\pi} + (\frac12 - \beta P) \ln \beta  \right) \,,\, \label{eq:mucl}
         a =  \frac{\partial \mu^{cl}}{\partial P } \,,
    \end{equation}
respectively. For a fixed $\beta$, $\mu^{cl}(P)$ is not a one-to-one function (see inset of Fig.~\ref{Fig1}), so there is an ambiguity in defining a thermal state by $\beta$ and $\mu^{cl}$, as in TBA. This is lifted by further requiring \begin{equation} a \ge 0 \,, \end{equation}
which selects the low-pressure branch. We observe that the latter solution is systematically selected by the solution (by numerical iteration) of the classical TBA equation \ref{eq:class_yy}. Intuitively, this is because a state with $a < 0$ is inaccessible from the asymptotic Bethe Ansatz that we assumed initially (see Section~\ref{sec:BB} above).

To sample a two reservoir initial condition with $2N$ particles, we generate $p_1, \dots, p_N, r_1, \dots, r_{N}$ using Eq.~\ref{eq:Z} with $\beta_L, \mu_L$, and $p_{N+1}, \dots, p_{2N}, r_{N+1}, \dots, r_{2N-1}$ with $\beta_R, \mu_R$. The overall translation freedom is fixed by requiring $(x_N + x_{N+1})/2 = 0$.
A result of the comparison is shown in Fig.~\ref{Fig1}. We find an excellent agreement already at a moderate space-time scale, $x, t \sim 10^2$. At shorter times, the effect of higher-derivative corrections becomes visible, especially for the particle density.



\subsection{Lax matrix and quasiparticles}\label{sec:lax}
It is well-known that the equation of motion of the classical Toda lattice can be expressed in terms of a Lax pair~\cite{Flaschka}, as
\begin{eqnarray}
    \dot{L} = [B, L] \label{eq:Lax} \, .
\end{eqnarray}
For an open chain, $B$ and $L$ are the following tridiagonal matrices, which we shall regard as operators on the Hilbert space spanned by the orthonormal basis $\vert n \rangle, n = 1, \dots, N$:
\begin{eqnarray}
    L = \sum_{n=1}^N  b_n \vert n \rangle \langle n \vert  + 
    \sum_{n=1}^{N-1} a_n (\vert n \rangle \langle n + 1 \vert + 
    \vert n + 1\rangle \langle n  \vert) \,,\,  \label{eq:L} \\
    B =  \frac12 \sum_{n=1}^{N-1} a_n 
    (\vert n + 1\rangle \langle n  \vert - \vert n \rangle \langle n + 1 \vert ) \,.
   \end{eqnarray}
Here,
\begin{equation}
\label{eq:anbn} 
b_n = p_n \,,\, a_n = e^{-r_n/2}\, 
\end{equation}
are the Flaschka variables (up to a factor $2$). Eq.~\ref{eq:Lax} implies that the spectrum of the Lax matrix $L$ is conserved, as are the quasimomenta  in the kinetic theory. Can they be identified? Here we show that this is the case; moreover, the quasiparticles appear as eigenstates of the Lax matrix. 

\begin{figure}
    \centering
    \includegraphics[width=.65\textwidth]{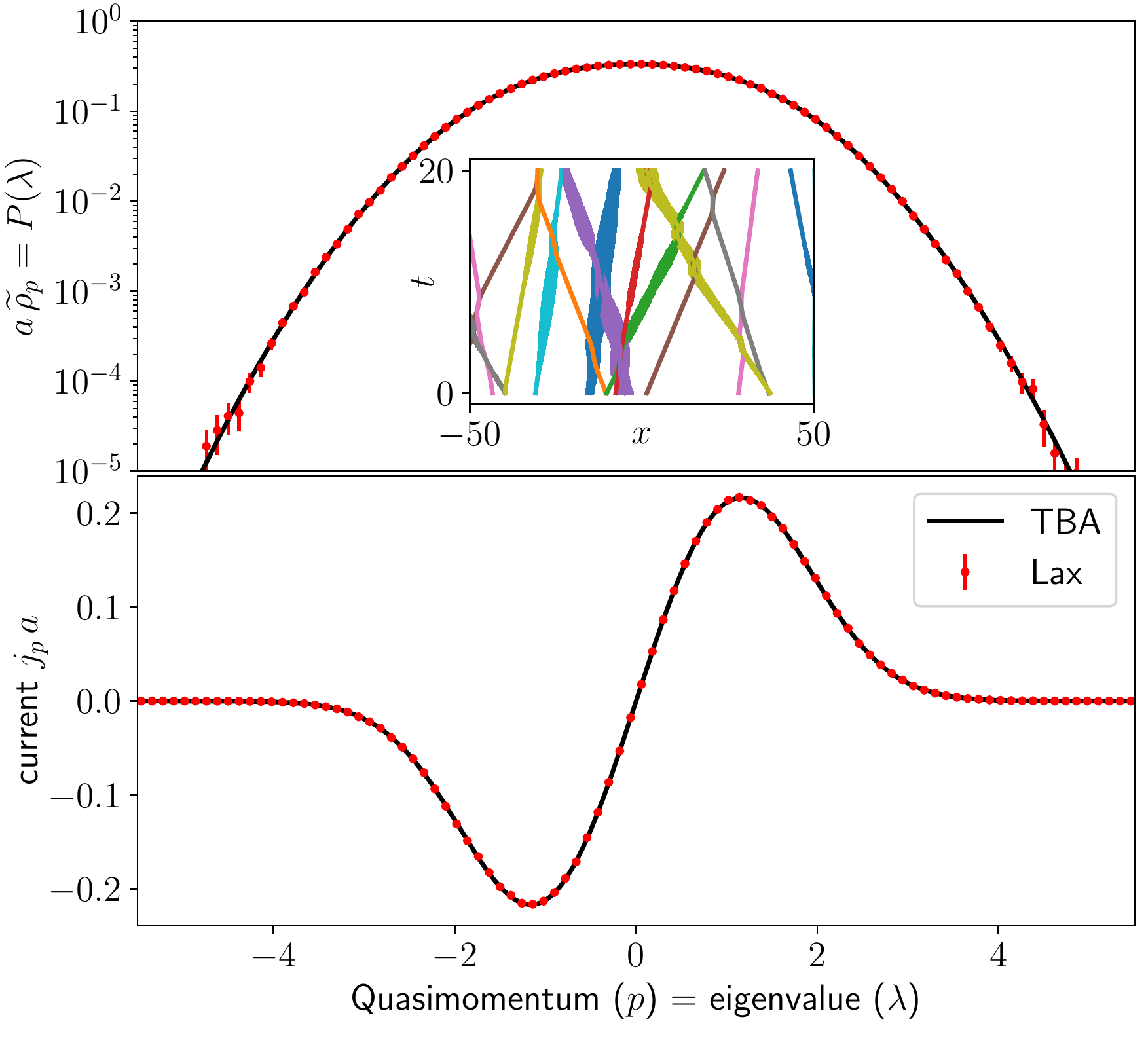}
    \caption{Main plots: Numerical verification of Eq.~\ref{eq:Pequalrho} (top panel) and Eq.~\ref{eq:jp} (bottom panel) for the thermal state $\beta = 1$, $\mu^{cl} = -1$ (average stretch $a=7.641$, pressure $P=0.138$). In both panels, the kinetic theory predictions are shown in curves, and the Lax matrix exact diagonalization results are shown in markers. The latter is averaged over $5 \times 10^6$ samples of size $N=2048$. The inset shows the time evolution of a few Lax eigenstates in an open chain of length $N=64$ from one thermal configuration. An eigenstate at a given time is drawn as a line centered around its expected position, with a width equal to its position uncertainty (standard deviation). Colors distinguish different eigenstates.}
    \label{fig:Lax}
\end{figure}

First, we compare the quasiparticle density $\rho(k)$ with the density of states $P(\lambda)$ of the Lax matrix for a thermal state at inverse temperature $\beta$ and pressure $P$. The former is determined by the TBA equation~\ref{eq:class_dos}. $P(\lambda)$ is the averaged density of states for the random Lax matrix ensemble whose matrix elements $a_n, b_n$ are obtained from the thermal distribution~\ref{eq:Z} (via Eq.~\ref{eq:anbn}). In terms of the Green's function, we have
\begin{equation}
    P(\lambda) =  \frac{1}{i \pi} \left.\overline{  \frac1N\mathrm{Tr}[G(\lambda - i\epsilon)] }\right|_{\epsilon \to 0_+}\,,\, G(z) := (z - L)^{-1} \,.
\end{equation}
where $\overline{[\dots]}$ denotes the ensemble average. Note that $L$ is Hermitian and has a real spectrum. We computed $P(\lambda)$ by exact diagonalization for a few thermal states, and found that it agrees perfectly with $\widetilde{\rho}_p$ (this depends on the normalization of Eq.~\ref{eq:anbn}), up to a normalization constant given by the average stretch:
\begin{equation}
    P(\lambda) = a \, \widetilde\rho_{p} \,,\, \mathrm{where } \, \lambda = p \,. \label{eq:Pequalrho}
\end{equation}
One example is shown in Fig.~\ref{fig:Lax} (top panel). 

We next turn to the quasiparticle currents, given by
\begin{equation} j_p = v_p[\rho]\widetilde\rho_p = \frac{1}{2\pi} \widetilde\theta_p {\epsilon^{cl}_p}'  \end{equation}  
in the kinetic theory. To obtain their (conjectural) expression in terms of the Lax matrix, consider an eigenstate $\vert \lambda \rangle$ of $L$ with eigenvalue $\lambda$. It evolves under
\begin{eqnarray}
    \frac{d}{dt}\vert \lambda \rangle = B \vert \lambda \rangle \,. \label{eq:inverse}
\end{eqnarray}
We want to interpret this eigenstate as a quasiparticle. Its position is given by an expectation value
\begin{equation} 
x(\lambda) = \sum_{n} \langle \lambda \vert n \rangle x_n \langle n \vert \lambda \rangle \,,\, x_n = r_{n-1} + r_{n-2} + \dots + r_1 + x_1 \,,\end{equation}
which is not well-defined in terms of the Flaschka variables (because of the overall displacement). However, the time derivative of $x(\lambda)$ can be calculated without ambiguity using Eq.~\ref{eq:inverse}:
\begin{eqnarray}
    \dot{x}(\lambda) =  \langle \lambda \vert V \vert \lambda \rangle \,,\, \\
    V := \sum_n  \left[ b_n \vert n \rangle \langle n \vert 
     +  \frac12 a_n \ln a_n ( \vert n \rangle \langle n + 1 \vert + \vert n + 1 \rangle \langle n \vert)   \right] \,.
\end{eqnarray}
Summing over eigenstates, we have
\begin{equation}
    j_{p} = \frac{1}{i \pi} \left.\overline{  \frac1N \mathrm{Tr}[ V  G(p - i\epsilon)] }\right|_{\epsilon \to 0_+} \,. \label{eq:jp}
\end{equation}
We numerically evaluate the RHS for thermal states by exact diagonalization and compare it to the kinetic theory prediction, obtaining again an excellent agreement, see Fig.~\ref{fig:Lax} (bottom panel). 

The identification of quasiparticles with Lax eigenstates (which may be called a Bethe-Lax correspondence) is conceptually appealing: it explains why the former can be regarded as point-like objects at the level of kinetic theory. While this is straightforward in simple systems such as hard rods (a quasiparticle is literally a hard rod, modulo permutations), it is not \textit{a priori} clear for the classical Toda lattice. An intuitive justification at non-zero temperature is as follows: since $L$ is a tridiagonal matrix with random elements of order unity, it can be viewed as a one-dimensional, disordered tight-binding Hamiltonian. It is then well-known that its eigenstates are all Anderson localized~\cite{RevModPhys.57.287}. Therefore, the quasiparticles have a position uncertainty of order unity, and can be treated as points at the hydrodynamic scale. This is illustrated in Fig.~\ref{fig:Lax}, inset, where we show a space-time snapshot of Toda lattice evolution, from the viewpoint of scattering quasiparticles.

\section{Conclusion}\label{sec:discussion}
We have studied the kinetic theory of the quantum and classical Toda lattices, from the perspective of an interacting gas of particles on a line. While the quantum kinetic theory was obtained from TBA in a manner that is straightforward at this point, and the classical kinetic theory can be obtained directly~\cite{doyon}, our analysis of the semiclassical ($\hbar \to 0$) limit is physically instructive and allows us to quantify the size of quantum corrections through a systematic expansion in $\hbar$; this shows in particular that the classical limit is non-singular. This may not be surprising given that the quantum Toda lattice can be viewed as an infinite-spin limit of $SU(2)$ integrable spin chains~\cite{hikami96toda}, and so is already ``classical'' in some sense. 
Here, we found a hydrodynamical consequence of this fact, in terms of the semiclassical expansion of the phase shift (Eq.~\ref{eq:K_expansion}): at any order in $\hbar$, the quantum corrections only affect scatterings in the classically static limit. We suspect that this is a general phenomenon in the semiclassical hydrodynamics of integrable systems. For instance, this holds for the Lieb-Liniger model, for which $K^{cl}_{p,p'} = 0$ is moreover trivial. It will be interesting to analyze the implications of this for the dynamics of the trapped $\delta$-Bose gas, as was realized in the famous ``Quantum Newton's Cradle'' experiment~\cite{cradle}.

Our numerical studies validate the classical kinetic theory by a direct comparison with microscopic non-equilibrium dynamics, as well as justifying the underlying equations of state from the viewpoint of a Bethe-Lax correspondence. This relates predictions from kinetic theory with mathematical conjectures about certain tridiagonal matrix ensembles, which are closely related to the $\beta$-random matrices~\cite{rmt}. This relation was recently exploited~\cite{spohn} in order to construct the generalized Gibbs ensemble of the classical Toda lattice from a random-matrix point of view. A similar approach might be useful in developing a kinetic theory for the Calogero-Sutherland-Moser systems, which admit a similar Lax-pair formulation. It should be clear that mathematically, the Bethe-Lax correspondence is nothing more than the inverse scattering method. However, in the context of the hydrodynamics of integrable systems, it has not previously been realized in a concrete way. Doing so for quantum integrable systems (including the quantum Toda lattice~\cite{Siddharthan1996}) might significantly elucidate the connection between kinetic theory and microscopic quantum dynamics, and represents an exciting challenge for future study.

\vspace{.3cm}
\textit{Acknowledgments.} We thank Herbert Spohn for illuminating discussions and for sharing his results prior to their publication in \cite{spohn}. We acknowledge support from the DARPA DRINQS program (VBB), the ERC synergy Grant UQUAM, DOE grant DE-SC0019380 (XC), and the Center for Novel Pathways to Quantum Coherence in Materials, an Energy Frontier Research Center funded by the U.S. Department of Energy, Office of Science, Basic Energy Sciences, and a Simons Investigatorship (JEM). Numerical computations are carried out on the cluster of Laboratoire de Physique Th{\'e}orique et Mod{\`e}les Statistiques (Orsay, France).

\vspace{.2cm}
\bibliography{Todabib}

\end{document}